\def \gapx {\lower 2pt \hbox{ $\buildrel>\over{\scriptstyle{\sim}}$ }}
\def \lapx {\lower 2pt \hbox{ $\buildrel<\over{\scriptstyle{\sim}}$ }}
\def \etal {{\it et al. \/}}
\def \guth {Guth 1981}
\def \cobe {Mather \etal 1990; Hogan 1990}
\def \COBE {Smoot \etal 1993; Wright \etal 1993}
\def \lstr {Lynden--Bell \etal 1988; Clowes and Campusano 1991}
\def \bc {Bonnor and Chamorro 1990; 1991}
\def \lem {Lema\^{\i}tre 1931; 1933}
\def \tol {Tolman 1934} 
\def \bon {Bondi 1947}
\def \bonnor {Bonnor 1972; 1974}
\def \pacz {Paczy\'{n}ski and Piran 1990}
\def \motat {Moffat and Tatarski 1992}
\def \grwall {Geller and Huchra 1989}
\def \iras {Efstathiou \etal 1990; Saunders \etal 1991}
\def \broadhurst {Broadhurst \etal 1990}
\def \sancisi {Sancisi and van Albada 1987}
\def \surveys {Maddox \etal 1990; Tyson 1988; Heydon-Dumbleton
\etal 1989; Lilly 1993}
\def \nmotat {Moffat and Tatarski 1994}
\def \cosmoloc {Weinberg 1972; Kolb and Turner 1990}
\def \peebs {Peebles 1993}
\def \sandage {Sandage 1972}
\def \krasb {Krasi\'{n}ski 1994}
\def \sato {Sato 1984; Suto \etal 1984a, 1984b}
\def \satos {Sato 1984}
\def \cham {Chamorro 1991}
\def \lilobs {Lilly 1994}
\def \vadas {Vadas 1993}

\documentstyle[12pt,aaspp,subeqn]{article}
\begin{document}
\begin{titlepage}

\title{COSMOLOGICAL OBSERVATIONS IN A LOCAL VOID\\}
\vspace{.75in}
\author{J. W. Moffat $^{\dag}$ \& D. C. Tatarski $^{\ddag}$\\}
\affil{Department of Physics\\  University of Toronto\\
Toronto, Ontario M5S 1A7, Canada\\}
\vspace{.5in}
\begin{abstract}
A local void in the globally Friedmann-Robertson-Walker (FRW)
cosmological model is studied. The inhomogeneity is described using
the Lema\^{\i}tre-Tolman-Bondi (LTB) solution with the spherically
symmetric matter distribution based on the faint galaxies number
counts. We investigate the effects this has on the measurement of
the Hubble constant and the redshift--luminosity distance relation
for moderately and very distant objects ($z \approx 0.1$ and more).
The results, while fully compatible with cosmological observations, indicate
that if we happened to live in such a void, but insisted on interpreting
cosmological observations through the FRW model, we could get a few
unexpected results. For example the Hubble constant measurement could
give results depending on the separation of the source and the observer.
\end{abstract}

\vspace{.5in}
\dag: moffat@medb.physics.utoronto.ca

\ddag: tatarski@medb.physics.utoronto.ca

\vspace{.25in}
{\bf \large UTPT-94-19}

\end{titlepage}

\section{INTRODUCTION}

It seems, particularly after the introduction of the inflationary paradigm
(\guth), that the isotropic and homogeneous Friedmann-Robertson-Walker
(FRW)  cosmological models are best suited for the description of the global
structure and the evolution of the universe. However, a similar statement is
not necessarily true when cosmologically moderate scales are thought
of.

There exists direct observational evidence in favour of the large scale
isotropy of the observed universe, namely, the COBE data confirming a high
degree of isotropy of the cosmic microwave background radiation (CMBR)
(\cobe; \COBE). However, the observational basis for the other standard
assumption made in the FRW cosmology, the homogeneity, is weaker.
There exists observational evidence in favour of larger and larger structures
(e.g. \lstr; \grwall).

We think that there exists sufficient observational evidence (discussed later
in this paper) to support a conjecture that we may live in a relatively large
underdense region embedded in a globally FRW universe. Exploring
observational properties of such a model is the aim of the present paper.

The simple model presented in this paper is restricted by the two physical
demands we impose on it. Firstly, to be consistent with the CMBR isotropy
it has to be spherically symmetric. Secondly, the model should be very
similar to an FRW one at the beginning of the expansion, but become
observably different at later times. In this manner, we could retain the
accomplishments of the FRW cosmology in dealing with early epochs,
while gaining new freedom in modelling the more recent universe.

The work on modelling voids of the Lema\^{\i}tre-Tolman-Bondi type
\footnote{A cosmological solution spherically symmetric about one point
was first proposed by Lema\^{\i}tre (\lem). However, it is usually called
the Tolman--Bondi solution (\tol, \bon).} in the expanding FRW universe
has been extensive. Main results and references will be briefly discussed
later. (An excellent review of the inhomogeneous cosmology exists: \krasb.)
Nevertheless, the observable consequences of such a model have seldom
been studied (e.g. \pacz; \motat).

In the following section, we briefly describe the LTB model. Section
\ref{void} consists of a concise review of the main results in modelling LTB
voids, a discussion of the observational background and the simple model
of a local void presented here. The description of our results of numerical
calculations is contained in Section \ref{resnum} The closing section
contains a discussion and conclusions.

Throughout this paper we use units in which $G=c=1$, unless stated
otherwise. Moreover, we choose the cosmological constant $\Lambda=0$.

\section{MODEL} \label{model}

First, for the sake of notational clarity, let us recall the FRW line element:
\begin{equation} \label{frwlinel}
 ds^{2}=dt^{2}-{a}^{2}(t)\left[ \frac{dr^{2}}{1-kr^{2}} +         
 r^{2}d\Omega^{2} \right],
\end{equation}
with $d\Omega^{2}={d\theta}^{2}+{\mbox{sin}}^{2}\theta{d\phi}^{2}$.

Now, let us consider a Lema\^{\i}tre-Tolman--Bondi (\tol, \bon) model for
a spherically symmetric inhomogeneous universe filled with dust. The line
element in comoving coordinates can be written as: 
\begin{equation} \label{linel}
 ds^{2}=dt^{2}-R^{\prime 2}(t,r)f^{-2}dr^{2}-R^{2}(t,r)d\Omega^{2},
\end{equation}
where $f$ is an arbitrary function of $r$ only, and the field equations
demand that $R(t,r)$ satisfies:
\begin{equation} \label{F}
 2R\dot{R}^{2}+2R(1-f^{2})=F(r),
\end{equation}
with $F$ being an arbitrary function of class $C^{2}$, $\dot{R}=\partial R /
\partial t$ and $R^{\prime}=\partial R / \partial r$. We have three distinct
solutions depending on whether $f^{2}<1$, $=1$, $>1$ and they
correspond to elliptic (closed), parabolic (flat) and hyperbolic (open) cases,
respectively.

The proper density can be expressed as:
\begin{equation} \label{dens}
 \rho=\frac{F^{\prime}}{16\pi R^{\prime} R^{2}}.
\end{equation}

Whatever the curvature, the total mass within comoving radius $r$
is:
\begin{equation} \label{mass}
M(r)= \frac{1}{4} \int_{0}^{r} dr f^{-1} F^{\prime} =4 \pi
\int_{0}^{r} dr \rho f^{-1} R^{\prime} R^{2},
\end{equation}
so that
\[
 M^{\prime} (r) = \frac{dM}{dr} = 4 \pi \rho f^{-1} R^{\prime}
R^{2}.
\]
Also for $\rho > 0$ everywhere we have $F^{\prime} >0$ and
$R^{\prime} >0$ so that in the non-singular part of the model $R>0$
except for $r=0$ and $F(r)$ is non-negative and monotonically
increasing for $r \geq 0$. This could be used to define the new
radial coordinate $\bar{r}^{3}=M(r)$ and find the parametric
solutions for the rate of expansion.

In the flat (parabolic) case $f^{2}=1$, we have
\begin{equation}
R = \frac{1}{2}{\left(9F\right)}^{1/3}{\left(t+\beta\right)}^{2/3},
\end{equation}
with $\beta(r)$ being an arbitrary function of class $C^{2}$ for
all $r$. After the change of coordinates $R(t,\bar{r}) = \bar{r}
{\left(t+\beta(\bar{r})\right)}^{2/3}$, the metric becomes:
\begin{equation} \label{metric}
 ds^{2}=dt^{2}-{\left(t+ \beta \right)}^{4/3} \left( Y^{2}        
  dr^{2}+r^{2}d\Omega^{2} \right),
\end{equation}
where
\begin{equation}
 Y= 1 + \frac{2 r {\beta}^{\prime}}{3 \left(t+ \beta \right)},
\end{equation}
and from (\ref{dens}) the density is given by
\begin{equation} \label{densb}
 \rho = \frac{1}{6 \pi {\left(t+ \beta \right)}^{2} Y} .
\end{equation}
Clearly, we have that ($t \rightarrow \infty$) the model tends to
the flat Einstein--de Sitter case.

For the closed and open cases the parametric solutions for the rate
of expansion can be written as (\bonnor):
\begin{subequations} \label{parneg}
\begin{equation}
 R=\frac{1}{4} F {\left( 1 - f^{2} \right)}^{-1} \left[1-\cos(v)
\right] ,\quad    f^{2}<1 ,
\end{equation}
\begin{equation}
 t+\beta= \frac{1}{4} F {\left( 1-f^{2} \right)}^{-3/2}\left[ v-
\sin(v) \right]   ,\quad f^{2}<1 ,
\end{equation}
\end{subequations}
and
\begin{subequations} \label{parpos}
\begin{equation} \label{parposR}
 R= \frac{1}{4} F {\left( f^{2}-1\right)}^{-1} \left[\cosh(v)-1
\right] ,\quad   f^{2}>1 ,
\end{equation}
\begin{equation} \label{parpost}
 t+\beta= \frac{1}{4} F {\left( f^{2}-1\right)}^{-3/2} \left[
\sinh(v)-v \right]   ,\quad f^{2}>1 ,
\end{equation}
\end{subequations}
with $\beta(r)$ being again a function of integration of class
$C^{2}$ and $v$ the parameter.

The flat case ($f^{2}=1$) has been extensively studied elsewhere (\motat).
The model depends on one arbitrary function $\beta (r)$ and could be
specified by assuming the density on some space-like hypersurface, say
$t=t_{0}$. However, specifying the density on the past light cone of the
observer is more appropriate.

The cases of interest to us, (\ref{parneg}) and (\ref{parpos}), correspond to
closed and open models, respectively.

Before we proceed (in the next section) to discuss the observational grounds
for modelling a local void, we need to amplify the discussion of the LTB
model by introducing basic features of the propagation of light. The high
degree of isotropy of the microwave background forces us to the conclusion
that we must be located close to the spatial centre of the local LTB void. In
our discussion, for the sake of simplicity, we place an observer at the centre
($t_{Ob}=t_{0} , r_{Ob}=0$).

The luminosity distance between an observer at the origin of our
coordinate system ($t_{0},0$) and the source at
($t_{e},r_{e},\theta_{e},\phi_{e}$) is (\bon):
\begin{equation} \label{lumdis}
 d_{L}={\left(\frac{{\cal L}}{4\pi{\cal F}}\right)}^{1/2}=R(t_{e},r_{e}){\left[1
 + z(t_{e},r_{e}) \right]}^2,
\end{equation}
where ${\cal L}$ is the absolute luminosity of the source (the
energy emitted per unit time in the source's rest frame), ${\cal
F}$ is the measured flux (the energy per unit time per unit area as
measured by the observer) and $z(t_{e},r_{e})$ is the redshift
(blueshift) for a light ray emitted at ($t_{e},r_{e}$) and observed
at ($t_{0},0$).

The light ray travelling inwards to the centre satisfies:
\[
 ds^{2}=dt^{2}-R^{\prime 2}(t,r)f^{-2}dr^{2}=0 ,\quad
d\theta=d\phi=0,
\]
and thus
\begin{equation} \label{zero}
 \frac{dt}{dr}=-R^{\prime}(t,r)/f(r),
\end{equation}
where the sign is determined by the fact that the light ray travels along the
{\em past} light cone (i.e. if $r_{e'}>r_{e''}$, then $t_{e'}<t_{e''}$).

Without getting into a detailed discussion, which can be found in \bon, or
\motat, let us state that if the equation of the light ray travelling along the
light cone is:
\begin{equation} \label{rays}
 t = T(r) ,
\end{equation}
using (\ref{zero}), we get the equation of a ray along the path:
\begin{equation} \label{lightcone}
 \frac{dT(r)}{dr}=-\frac{R^{\prime}}{f} [T(r),r] ,
\end{equation}
where 
\[
 {\dot{R}}^{\prime}[T(r),r] = 
 {\left. \frac{{\partial}^{2}R}{\partial t \partial r} \right     
 |}_{r,T(r)}   ={\left. \frac{\partial R^{\prime}}{\partial t}    
 \right |}_{r,T(r)} .
\]
The equation for the redshift considered as a function of $r$ along
the light cone is:
\begin{equation} \label{zred}
 \frac{dz}{dr}=(1+z){\dot{R}}^{\prime}[T(r),r] ,
\end{equation}
and the shift $z_{1}$ for a light ray travelling from ($t_{1},r_{1}$) to
($t_{0},0$) is:
\begin{equation} \label{grco}
  \ln (1+z_{1})  =  - \ln (1-a_{1}) - \int_{0}^{r_{1}} dr
\frac{M^{\prime}(r)}{r(1-a_{1})} ,
\end{equation}
where 
\[
a_{1}(r)=\dot{R}[T(r),r],
\]
and, in obtaining equation (\ref{grco}), we used (\ref{dens}) and
(\ref{mass}). Thus we have two contributions to the redshift. The
cosmological redshift due to expansion, described by the first term with
$a_{1}=\dot{R}$, and the gravitational shift due to the difference between
the potential energy per unit mass at the source and at the observer.
Obviously, in the homogeneous case ($M^{\prime}(r)=0$) there is no
gravitational shift.

\section{LOCAL VOID} \label{void}

If we restrict ourselves to spatial scales that have been well probed
observationally, i.e. up to a few hundred Mpc, the most striking feature of
the luminous matter distribution is the existence of large voids surrounded
by sheet-like structures containing galaxies (e.g. \grwall). These surveys (see
also \iras) give a typical size of the voids of the order 50--60 $h^{-1}$ Mpc.
There has also been some evidence (\broadhurst) --with less certainty-- for
the existence of larger underdense regions with characteristic sizes of about
130 $h^{-1}$ Mpc. Also, dynamical estimates of the FRW density
parameter ${\Omega}_{0}$ give very different results on different scales.
The observations of galactic halos on scales less than about 10 to 30
Mpc typically give (see e.g. \sancisi) ${\Omega}_{10-30} \simeq 0.2 \pm
0.1$. On the other hand, smoothing the observations over larger scales
($>20 \mbox{ Mpc}$, say $\sim 100 \mbox{ Mpc }$) indicates (e.g. \iras)
the existence of a less clustered component with a contribution
exceeding 0.2, and perhaps as high as ${\Omega}_{\sim 100} \simeq
0.8 \pm 0.2$.

At the same time, the large scale galaxy surveys (some of the recent
literature is given in \surveys) firmly indicate a considerable excess in
the number--magnitude counts for faint galaxies relative to predictions of
homogeneous, ``no-evolution'' models. This excess could be the result of  a
non-standard galactic evolution or could be caused by rather exotic FRW
cosmology (i.e. the deceleration parameter $q_{0} \ll 0.5$ or a non-zero
cosmological constant $\Lambda$). However, it can also be treated as an
observational indication of a very large (on the scale of the redshift $z
\approx 0.5$) void. In the following, we choose this latter option in
interpreting the faint galaxies number counts and model the density
distribution of a local void accordingly. (A somewhat similar model of a
gaussian void as a function of the comoving coordinate can be found in
\nmotat.)

\begin{figure}[ht]
\vspace{3.4in} \relax \noindent \relax
\includegraphics{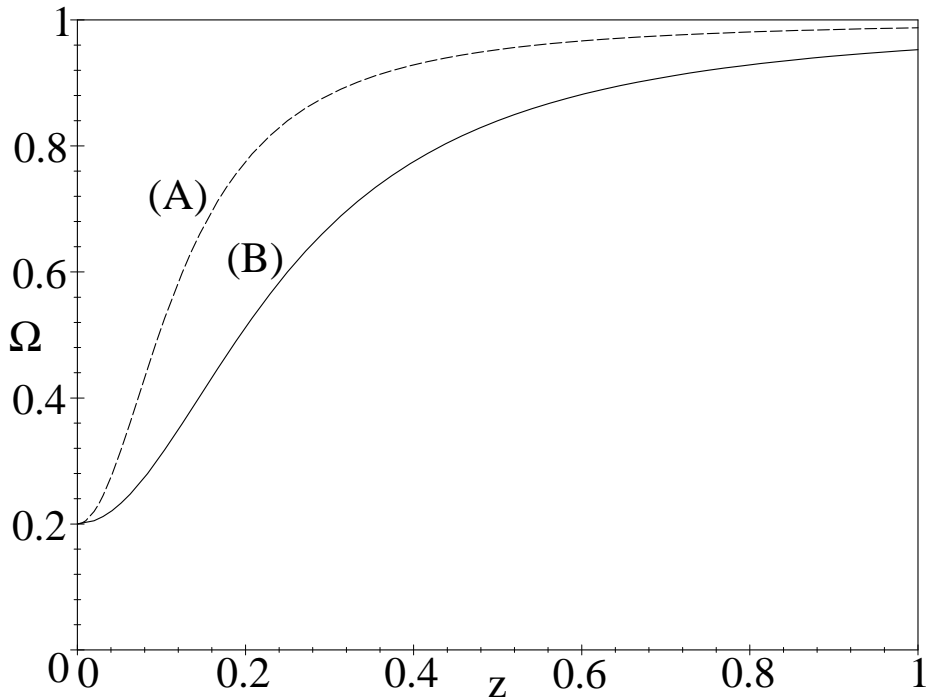}
\caption{{\tenrm\baselineskip=10pt The density distributions
$\Omega_v$ as functions of the redshift $z$. The parametrization of the
cases (A) and (B) is given in the text. \label{f1}}}
\end{figure}

We study a void with the central density equal to that of an FRW model
with the density parameter $\Omega_{0}=0.2$, asymptotically approaching
the FRW model with $\Omega_{0}=1$. Since cosmological observations
are done by detecting some form of electromagnetic radiation, the mapping
obtained from them describes the density along the light cone. We describe
the density distribution, as a function of the redshift $z$, by:
\begin{equation} \label{densdistr}
\Omega_v(z)=\frac{\Omega_{min}+{(z/a)}^2}{\Omega_{max}+{(z/a)}^2}.
\end{equation}
The choice of a rational function $\Omega_v(z)$ (as opposed to, say, a
gaussian distribution) assures that the differential equations that we solve
numerically (already quite complex) are not unnecessarily complicated
further. In the numerical calculations presented in the next section we used
the values $\Omega_{min}=0.2$, $\Omega_{max}=1$. The two density
distributions presented there are parametrized by (A) $a=0.125$ and (B)
$a=0.25$, and are depicted in Figure \ref{f1}.

In the following figures the results obtained for the (steeper) distribution (A)
will be represented by dashed lines, while the results for the density
distribution (B) will always be given as solid lines. Also, since we study
here the observations of the luminous matter, we present our results in the
range of up to $z \approx 6$. (An observation of an object with a redshift
in excess of this value would be hard to reconcile with standard FRW
cosmology.)

The final question we wish to address in this section is that of shell crossing
in the LTB models. The problem has been extensively studied (for a detailed
review and a complete list of references see \krasb).

As seen from (\ref{linel}) and (\ref{dens}) shell crossing can occur when
$R^{\prime}(t,r)=0$ for some $r=r_s$. This, in general, may lead to
$\lim_{r \rightarrow r_s} \rho = \infty$ and a shell crossing singularity.
Different shells $r=\mbox{const}$ collide and the comoving coordinates
become inadmissible. The singularity can be avoided, if the functions
$F^\prime(r)$ and $f(r)$ both have at $r=r_s$ zeros of the same order as
$R^\prime(t,r)$ has. This, however, is not the case in the present work (we
assume $F(r) \geq 0$ and $F^\prime(r) \geq 0$).

In the simple model presented here, shell crossing does occur. The physical
significance of this can be viewed from diametrically opposite positions.
One extreme is to regard a shell crossing singularity as unphysical and try
to exclude it by assuming special initial conditions, as was the strategy in
the work of Bonnor and Chamorro (\bc; \cham). However, this leads to
difficulties with adjusting the model to observational data. In short: either
the average density of the void is not appreciably lower than the density of
the background, or the matter in the void is much older than outside. The
second approach, best represented by Sato and collaborators (see e.g.
\sato), is to treat the shell crossing singularity as a physical entity. In this
approach, shell crossings --numerically modelled as surface layers of
matter-- are interpreted as rims of the expanding voids and correspond to
the aforementioned sheet-like structures surrounding observed voids.

An observationally acceptable LTB void, with the density distribution similar
to that depicted in Figure \ref{f1}, is too deep to escape shell crossing and,
at the same time, allow for the simultaneous Big Bang throughout the whole
space. Since we see the latter property as very important in order to be able
to incorporate the accomplishments of the FRW cosmology in dealing with
epochs preceding the matter dominated era, we choose to accept shell
crossing as a price. The more so, that --as will be shown in the next
section-- the distribution (\ref{densdistr}), even though chosen on purely
observational grounds, leaves us with a bonus: the shell crossing happens in
the distant future. This, in our opinion, indicates that the LTB void studied
here is applicable to the matter dominated era. (The fact that the LTB model
does not allow for pressure is not perilous from our point of view. After all,
the universe has been matter dominated since $z \approx 10^4$. The
model is applicable to this era and may be thought of as a continuation of
an earlier --very nearly-- FRW model, provided its density contrast at the
time of decoupling, $t=t_{dc}$, is $\rho(t_{dc},r) / \rho_{FRW}(t_{dc})
\approx1$.)

\section{RESULTS} \label{resnum}

In general, an LTB model depends on three arbitrary functions (see section
\ref{model}), $F(r), \beta(r)$ and $f(r)$. Since $F(r)$ can be interpreted as
twice the effective gravitational mass within comoving radius $r$
(\bon), then, in accordance with the discussion following (\ref{mass}),
assuming its form is equivalent to a coordinate choice. In our calculations
we used $F(r)=4r^3$. The second function, $\beta(r)$, sets the initial
singularity hypersurface of the model. Since we want the region of highest
density in our model to be fully equivalent to the FRW universe, we set
$\beta(r)= \beta_0= \mbox{const}$, thereby assuming a universally
simultaneous Big Bang. In doing so, we give up a very important feature of
an LTB model: an extra (with respect to FRW) degree of freedom that would
allow the age of the universe to be position dependent. However, we study
the following two cases (both with the simultaneous Big Bang
hypersurface): the universe whose age is equal to that of the FRW critical
($\Omega_0 = 1$) case and the universe with the age equal to that of the
FRW $\Omega_0 = 0.2$ one. The parametrization of the cases is,
respectively, (I) $\beta_0 = \beta_I = 0$ and (II) $\beta_0= \beta_{II} =
1.23606$, where --in both cases-- we set the time coordinate of constant
time hypersurface ``now'' so that it is equal to the age of the universe
$t_{0}=1$ in the FRW model with $\Omega_{0}=1$. The third
(``curvature'') function, $f(r)$, is an unknown to be solved for in our
calculations. Since we are modelling an underdense comoving void in an
FRW universe, we choose the LTB hyperbolic ($f^2>1$) case.

In a manner similar to that employed in \motat, we assume that since
all cosmological observations are necessarily done by detecting some form
of electromagnetic radiation, the solution should progress along the light
cone. The final set of equations we solve consists of the equations
(\ref{lightcone}), (\ref{zred}) and the equation describing the ratio of the
local density (\ref{dens}) to the density of the FRW universe through the
relation (\ref{densdistr}) taken along the past light cone:
\begin{equation} \label{findens}
\frac{\rho[T(r),r]}{\rho_{FRW}[T(r)]}=\frac{3F^{\prime}(r){[T(r)+\beta_0]}^2}
{8 R^{\prime}[T(r),r] R^{2}[T(r),r]}=\Omega_{v}[z(r)].
\end{equation}
The final form of the equations is quite unwieldy and we do not present
them here.  Since $T(r)$ (the time of emission $t_{e}$ of a light ray
observed at $r=0$ at $t_{0}$) is now given by (\ref{parpost}) and $z(r)$ is
governed by (\ref{zred}), the functions to be solved for are $f(r), z(r)$ and
$v(r)$, where the parameter $v$ becomes the function of position.

The initial conditions for the integration have to be set at $r \neq 0$, since
the analytic expressions (\ref{parpos}) are singular at $r=0$, where $f^2=1$
(we have a flat $\Omega_{0}=0.2$ FRW universe there). We assume that
for the initial radius $r_i \ll 1$ (we use dimensionless radius and time in
the calculations) corresponding $z_i$ and $t_i$ are given by their standard
FRW values. Then $z(r_i), v(r_i)$ and $f(r_i)$ can be obtained from
(\ref{parpost}), (\ref{lightcone}) and (\ref{zred}).

Once the equations have been numerically integrated (in all cases the
accuracy of our numerical procedure was $10^{- 6}$ or better
\footnote{We do not elaborate here on the intricacies of the numerical
procedure. A good discussion on the technical aspects of numerically
solving a somewhat similar problem involving an LTB model can be found
in \pacz.}) we use (\ref{parpost}) to obtain $t_{e}$ for a given $r_{e}$. The
luminosity distance $d_{L}$ corresponding to this event is obtained with the
use of (\ref{lumdis}) and $z(t)$ (useful in studying the cosmological time
scale) is given by the parametric relation $[ T(r), z(r) ]$.

\begin{figure}[ht]
\vspace{3.4in} \relax \noindent \relax
\includegraphics{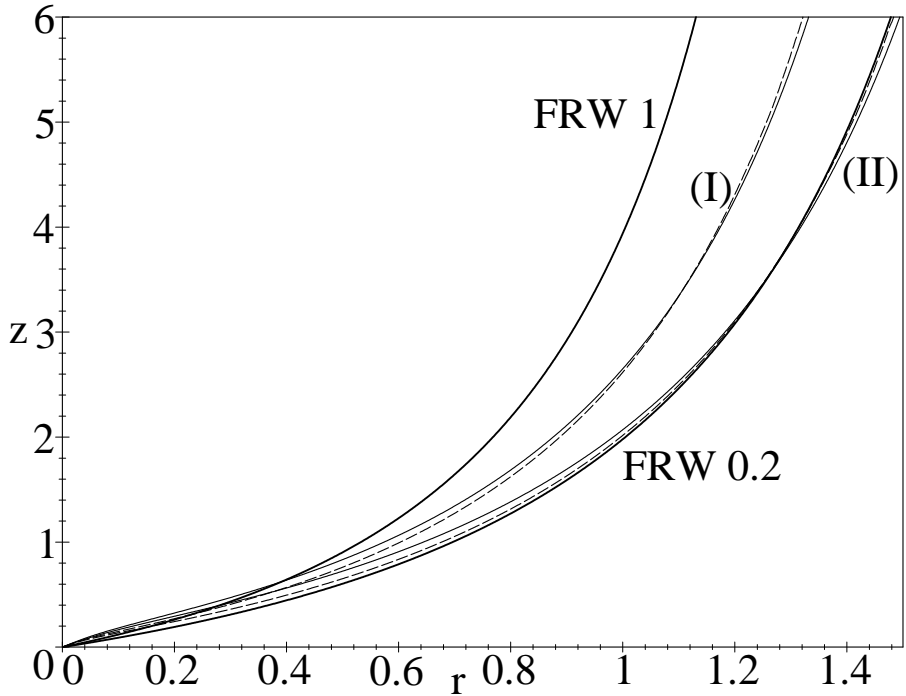}
\caption{{\tenrm\baselineskip=10pt The redshift $z$ as a function of the
comoving radius $r$. Roman numerals I and II denote the ``younger''
($\beta_0 = \beta_I$) and ``older'' ($\beta_0 = \beta_{II}$) solutions,
respectively. Dashed and solid lines correspondence to the two density
distributions of the void is the same as in Figure 1. FRW $\Omega=0.2$
and $\Omega=1$ solutions, respectively, are denoted ``FRW 0.2'' and
``FRW 1''.\label{f2}}}
\end{figure}

The results of our numerical calculations for $z(r)$, where $r$ is the
dimensionless comoving radius used in the calculations are depicted in
Figure \ref{f2}. The coordinate distance has no direct physical relevance,
but our units here are such that $r=1$ corresponds to $2997.95 h^{-1}$
Mpc, where $h$ is the usual coefficient in the observationally determined
value of the Hubble constant: $H_{0}=100 h \mbox{ km } {\mbox{s}}^{-1}
{\mbox{Mpc}^{-1}}$.

The behaviour of $z(r)$ is as intuitively expected. In the ``older'' case (II)
the redshift $z$ asymptotically converges to the FRW $\Omega_0=0.2$
behaviour. The additional gravitational shift caused by the mass distribution
of the void  is apparent on the scales comparable to that of the void ($z
\lapx 0.5$). On intermediate to large scales the effect of a cosmological
expansion is predominant. In the ``younger'' case (I), $z(r)$ --after an excess
over the FRW relation due to the gravitational shift induced by the void--
asymptotically tends to a limit that could, in accordance with the FRW
interpretation, correspond to the universe with the density parameter in the
range $\Omega \in (0.2;1)$. In both instances, the convergence to the FRW
relation is marginally faster for the cases with steeper density distributions.

However, one should not forget that the comoving distance is {\em
not} an observable, whereas the luminosity distance $d_L$ is. In principle,
provided we know its absolute luminosity $\cal{L}$, we can establish the
luminosity distance, defined by $4 \pi {d_L}^2=\cal{L} / \cal{F}$, by
measuring the energy flux $\cal{F}$ of an observed object (for a discussion
of usual caveats associated with so-called ``standard candles'' see e.g.
\cosmoloc; \peebs). It is the $z(d_L)$ relation that is observationally
relevant. Let us recall that the FRW equivalent of (\ref{lumdis}) is:
\begin{equation} \label{lumdisfrw}
{d_L}^2=a^2(t_0) {r_e}^2 {(1+z_e)}^2 .
\end{equation}

Before we probe the observable properties of the solution, let us investigate
the temporal evolution of the void. In doing so we seek answers to two
important questions. First, how soon in the further evolution of the void the
formation of a shell crossing singularity occurs. If this happens in the distant
enough future, then our model is applicable to a sufficiently long
cosmological epoch to be considered physical. Secondly, we want to verify
that at early times the density contrast $\rho_v/\rho_b \approx 1$, where
$\rho_v$ and $\rho_b$ are matter densities in the void and in the
background, in accordance with our conjecture that the model be very
similar to an FRW one at the beginning of the expansion.

In general, the earlier the formation time of the void, the faster the growth
of the density contrast. Also, the rate of growth is increased in the cases of
higher initial contrasts. (See \satos. A very thorough numerical study of the
evolution of general relativistic voids, including the LTB ones, can also be
found in \vadas.) In the study presented here, however, we did not assume
a density distribution on some hypersurface of constant time as initial
condition. The observationally based density distributions of Figure \ref{f1}
were given as functions of the redshift $z$ and since we numerically
integrated the equations along the light cone, we are not yet in the position
to examine the time evolution of the spatial distribution of the matter
density. However, equipped with the solution for $f(r)$ we can use
(\ref{parpost}) to obtain the values of the parameter $v_h(r)$ corresponding
to a given hypersurface of constant time $t=t_h$ throughout the space.
Then, using (\ref{dens}) we get the spatial matter density distributions for
the time $t_h$.

\begin{figure}[ht]
\vspace{3.4in} \relax \noindent \relax
\includegraphics{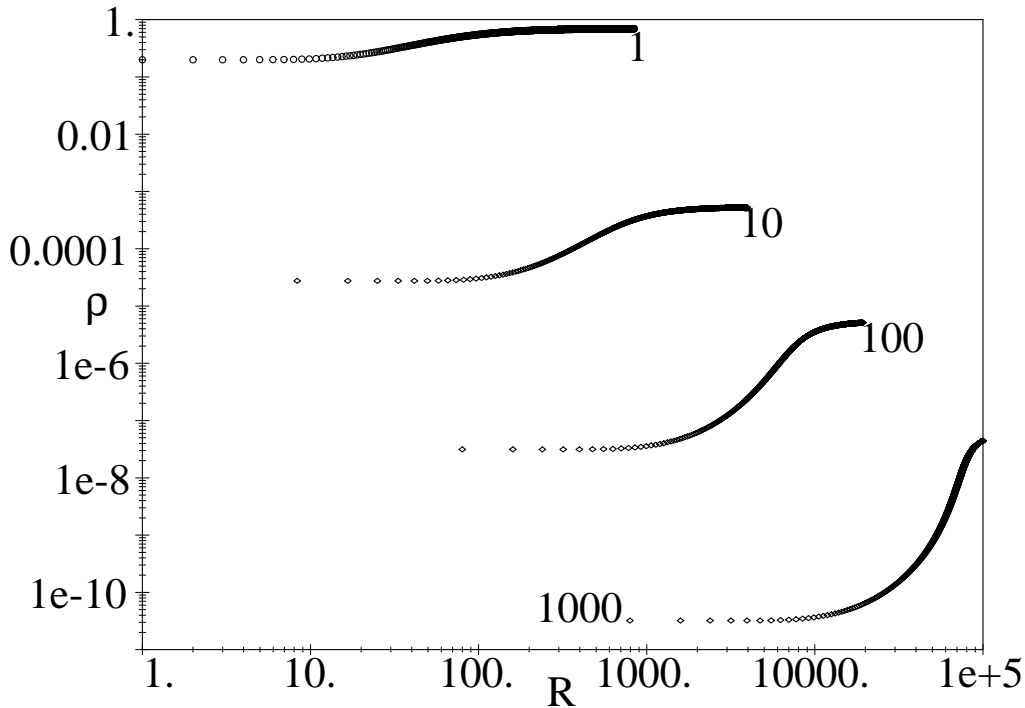}
\caption{{\tenrm\baselineskip=10pt The future evolution of the void (A) in
the case (I): $t_0+\beta_I=t_0=t_{FRW1}$. The matter density as a function
of $R(t_h,r)/R(t_0,r_i)$, where $r_i$ is the initial radius, is shown at times
$t_h=t_0, 10t_0, 100t_0 \mbox{ and } 1000t_0$. \label{f3}}}
\end{figure}

\begin{figure}[ht]
\vspace{3.4in} \relax \noindent \relax
\includegraphics{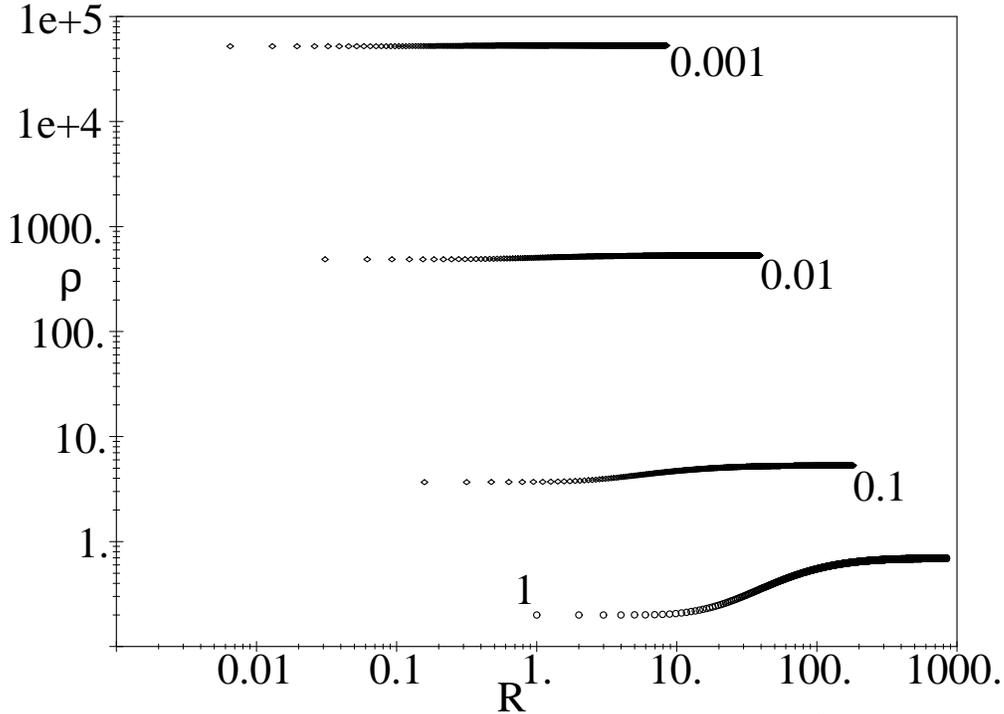}
\caption{{\tenrm\baselineskip=10pt The past evolution of the void (A) in
the case (I): $t_0+\beta_I=t_0=t_{FRW1}$. The matter density as a function
of $R(t_h,r)/R(t_0,r_i)$, where $r_i$ is the initial radius, is shown at times
$t_h=t_0, 0.1t_0, 0.01t_0 \mbox{ and } 0.001t_0$. \label{f4}}}
\end{figure}

The results (numerical solution points) of this calculation for the future and
past evolution of the steeper void (A) in the ``younger'' case (I) are
presented in Figures \ref{f3} and \ref{f4}, respectively. The density
distribution in this case evolves fastest and, as such, is the most
representative in estimating the time scales involved.

Figure \ref{f3} shows that the void expands outward increasing its density
contrast and forming a very steep ridge on its boundary. This (in
accordance with \satos; \vadas) leads to the eventual formation of a thin
dense shell and the shell crossing singularity. Of interest to us, however, is
the time scale of this process. Since the shell crossing occurs for $t
\gapx 1000t_0$, where $t_0$ is the present age of the universe, we
conclude that our model is applicable to a satisfactorily long cosmological
epoch.

At the same time, Figure \ref{f4} clearly indicates that our model satisfies
the requirement of similarity to the FRW one at early cosmological times. At
$t_h=0.001t_0$ the matter density is approximately constant everywhere.
For the case depicted in Figure \ref{f4} the density contrast is $\rho_v /
\rho_b \approx 0.982$ for $t_h=0.001t_0$ ($ \approx 0.996$ for $t_h =
10^{-4}t_0$). In all cases at $t/(t_0+\beta_0) \simeq 10^{-6}$ we have
$|\rho_v / \rho_b -1| \simeq 10^{-5}$. There is hardly a trace of the void's
presence.

\begin{figure}[th]
\vspace{3.3in} \relax \noindent \relax
\includegraphics{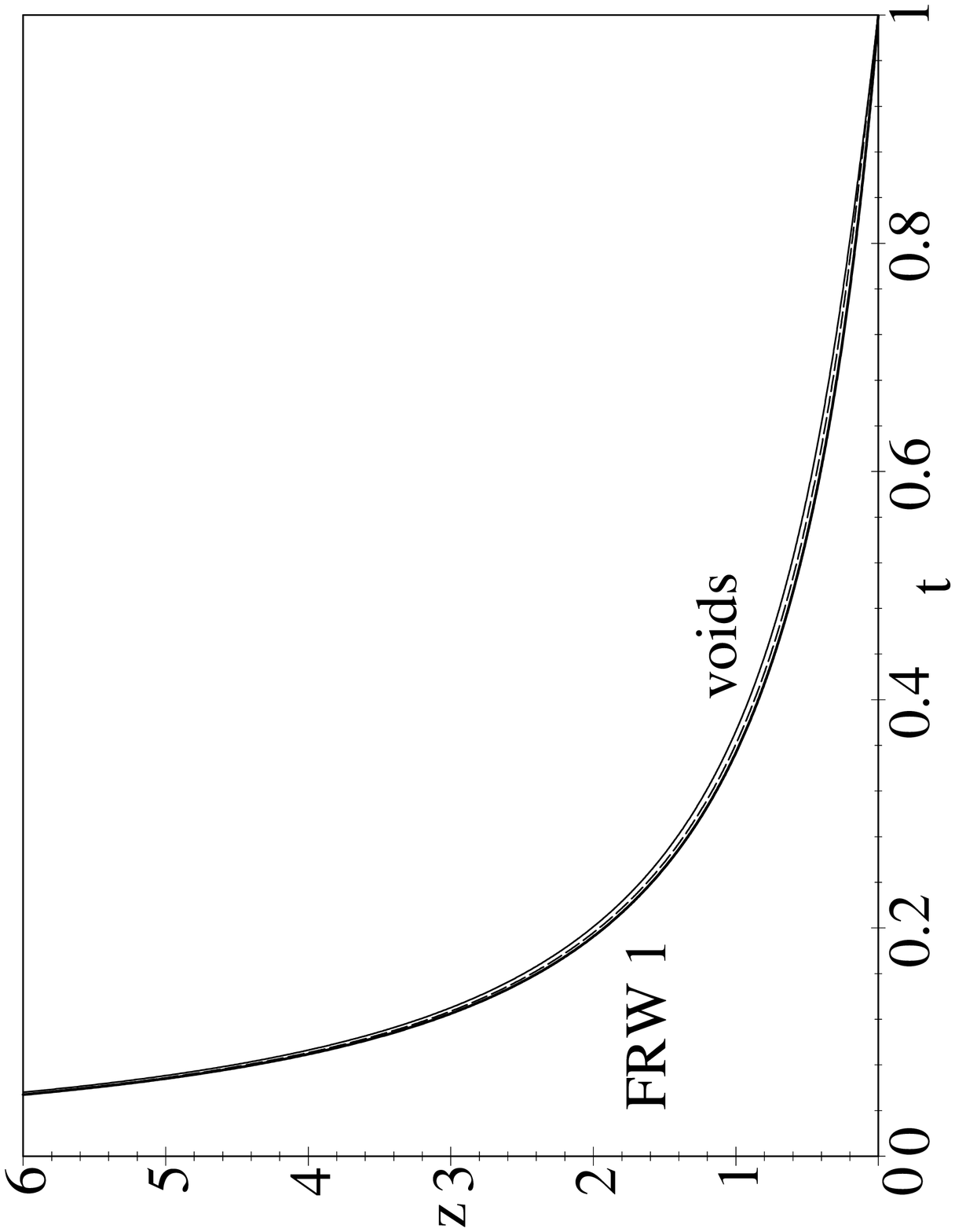}
\caption{{\tenrm\baselineskip=10pt The redshift $z$ as a function of the
cosmic time $t$ for the case (I): $t_0+\beta_I=t_{FRW1}$. ($t=1$
corresponds to the $t_{0}$ in the FRW critical case.) \label{f5}}}
\end{figure}

\begin{figure}[bh]
\vspace{3.4in} \relax \noindent \relax
\includegraphics{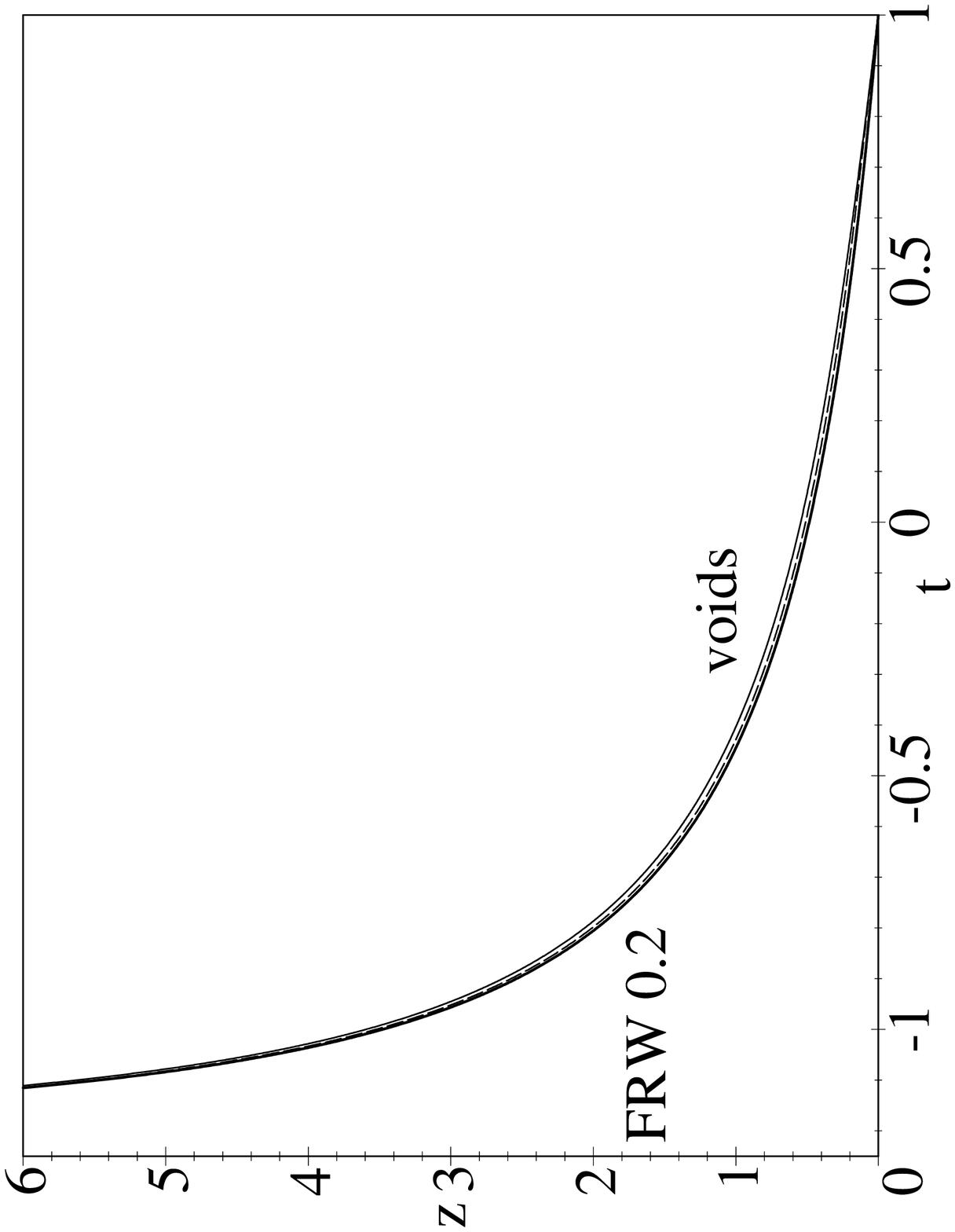}
\caption{{\tenrm\baselineskip=10pt The redshift $z$ as a function of the
cosmic time $t$ for the case (II): $t_0+\beta_{II}=t_{FRW0.2}$. ($t=1$
corresponds to the $t_{0}$ in the FRW critical case.) \label{f6}}}
\end{figure}

The $z(t)$ relations for the cases of different ``ages of the universe'' are
presented in Figures \ref{f5} and \ref{f6}. There is some divergence from the
FRW behaviour, but for early cosmological times the relations tend to their
FRW counterparts. This asymptotic behaviour is in accordance with our
assumption of a simultaneous big bang, $\beta(r) = \mbox{ const }$, and
with our setting the age of the universe in each case to be equal to that of
the corresponding FRW case. In both (I) and (II) cases objects with similar
redshifts are younger than their FRW counterparts (i.e. $t_{LTB} > t_{FRW}$,
when $z_{LTB}=z_{FRW}$), but not significantly. Those objects have
marginally more time to form and evolve than their FRW counterparts.

Let us now explore the $z(d_L)$ relations resulting from our model. Any
nonstandard cosmological model to be of interest must pass two tests when
confronted with observations. To be considered realistic, it is of utmost
importance that its predictions do not contradict the linearity of Hubble's
law $z=H d$, well established on small scales. At the same time, it should
be sufficiently different from the standard (FRW) models to allow for
significant reinterpretation of observational results. Let us now apply these
two tests to our model.

\begin{figure}[ht]
\vspace{3.4in} \relax \noindent \relax
\includegraphics{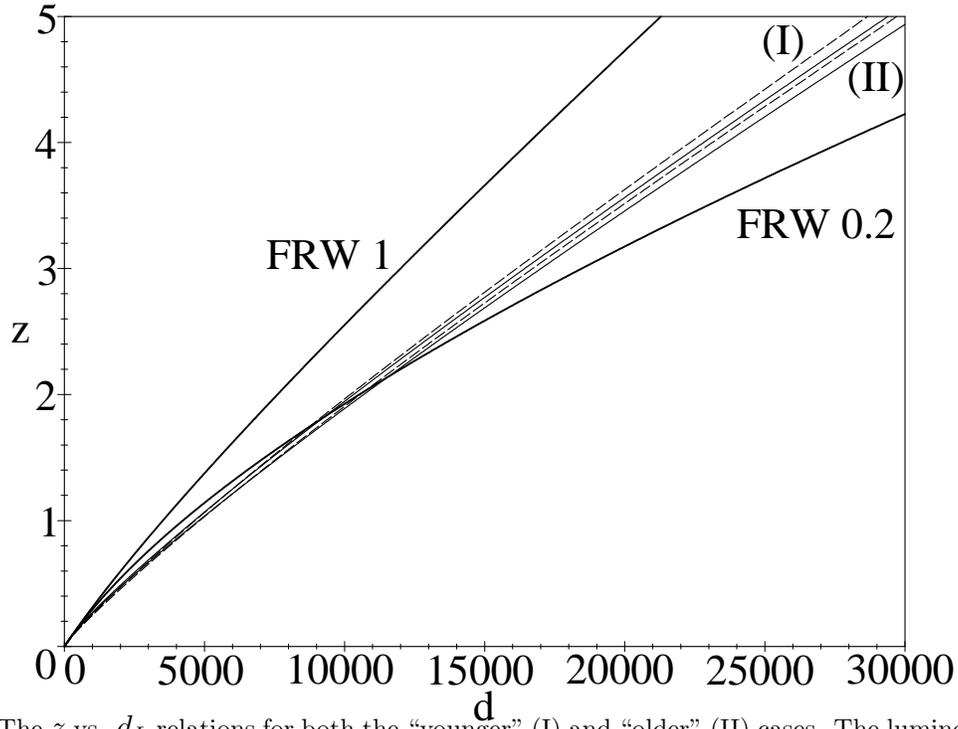}
\caption{{\tenrm\baselineskip=10pt The $z$ vs. $d_L$ relations for both
the ``younger'' (I) and ``older'' (II) cases. The luminosity distance $d_L$ is
given in Mpc. \label{f7}}}
\end{figure}

Figure \ref{f7} shows that (in all cases studied here) on large scales the LTB
``Hubble relation'' between the redshift $z$ and the luminosity distance
$d_L$ is reminiscent of the FRW relation for the universe with the density
parameter in the range $\Omega \in (0.2;1)$.

\begin{figure}[hb]
\vspace{3.4in} \relax \noindent \relax
\includegraphics{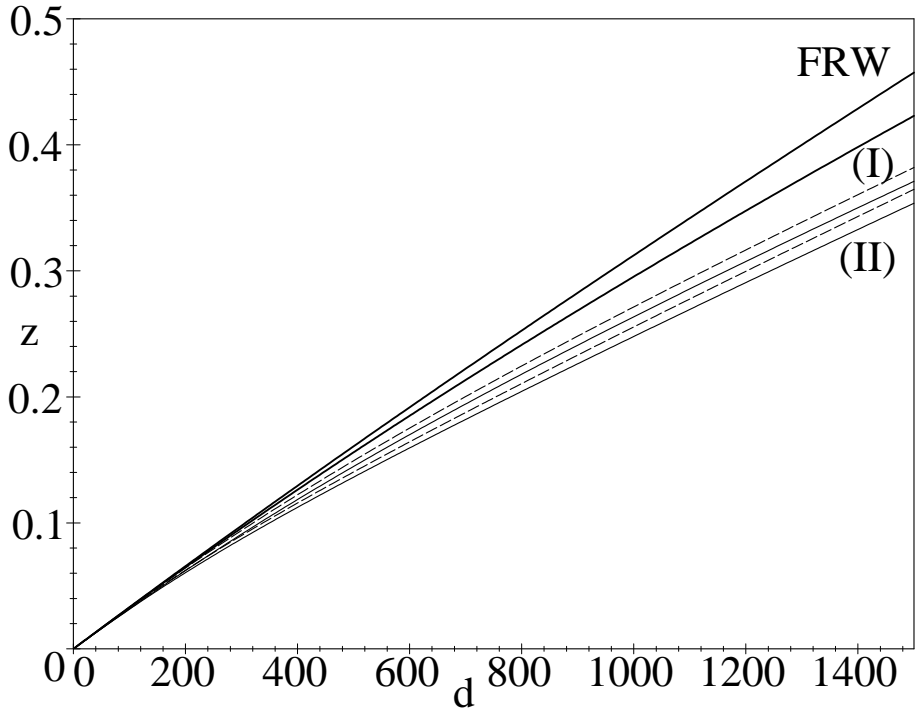}
\caption{{\tenrm\baselineskip=10pt The $z$ vs. $d_L$ relations for both
the ``younger'' (I) and ``older'' (II) cases for smaller scales. The luminosity
distance $d_L$ is given in Mpc. \label{f8}}}
\end{figure}

At the same time, inspection of Figure \ref{f8} shows that on small scales
very nearly linear ``Hubble diagrams'' are obtained for both types of LTB
voids and both time scales.
\newpage

\nopagebreak

\clearpage

\nopagebreak

\clearpage

If we recall (\sandage, a good overview can be found in \S5 of \peebs) that
the Hubble diagram, extended to $z \approx 0.5$, follows the Hubble's
law with the redshifts' scatter (assuming identical absolute magnitudes of
observed galaxies) of $\sim$ 20 \% around the theoretical prediction, then
we can claim that the redshift--distance (redshift--magnitude) relations
inferred from our model are within the observational bounds. Also,
inspection of Figure \ref{f8} shows that on small scales observationally
indistinguishable from linear ``Hubble diagrams'' are obtained. However, a
different value for the Hubble parameter (constant) is inferred, if we insist
on interpreting the results of cosmological observations through an FRW
model.

Also for larger observed redshifts (see e.g. the results of McCarthy's survey
of the state of the measurements of the infrared Hubble diagram for radio
galaxies, out to $z \approx 4$, in \S5 of \peebs) our results are well within
the observational scatter. Moreover, when the measurements of the Hubble
diagram are extended to large scales an interesting feature emerges. The
slope of the diagram at $z \lapx 1$ seems to be consistent with the linear
Hubble law. However, at $z \gapx 1$ a hint of curvature appears. In the
standard FRW interpretation, this is presumed to be the result of relativistic
corrections to the redshift--distance relation and of the fact that the
galaxies observed at high redshifts are seen as they were younger and very
likely more luminous than the galaxies observed at low redshifts.

The fact that the redshift is observed to increase with distance, at least
roughly in accordance with Hubble's law, for redshifts out to in excess
of unity, is usually heralded as one of the major observational confirmations
of standard FRW cosmology. It would be interesting to see how the
predictions of our model withstand the same observational test.

\begin{figure}[ht]
\vspace{3.4in} \relax \noindent \relax
\includegraphics{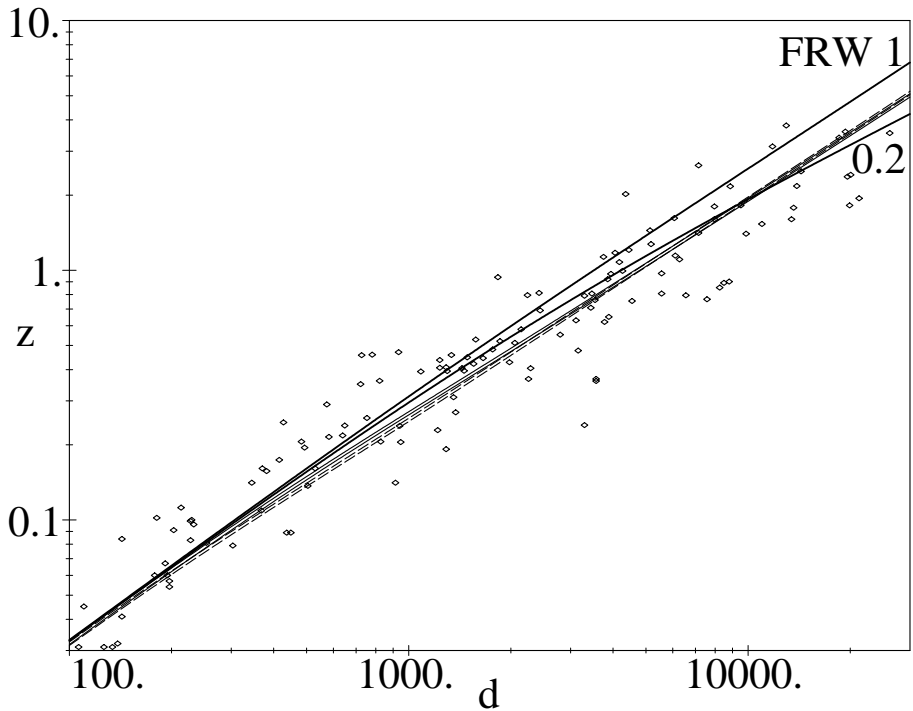}
\caption{{\tenrm\baselineskip=10pt The $\log(z)$ vs. $\log(d_L)$ relations.
Observations, denoted by $\diamond$, are adapted from \lilobs. The
luminosity distance $d_L$ is given in Mpc. \label{f9}}}
\end{figure}

Figure \ref{f9} depicts the logarithmic redshift--luminosity distance relations
for both LTB voids and both time scales. A survey of measurements
(adapted
from \lilobs) is also included. The $z(d_L)$ relations for the FRW cases with
$\Omega_0=1$ and $\Omega_0 = 0.2$ are presented for comparison.

It is now apparent that the redshift--luminosity distance relations inferred
from our model are in agreement with observations. For small redshifts, say
up to $z \approx 0.1$, they are observationally indistinguishable from the
FRW results. For intermediate redshifts --in the range, say, $z \in (0.1;1)$--,
where the gravitational shift contribution to the overall redshift causes the
most manifest departure from the FRW relations, our results remain well
within the observational scatter. In the range $z \gapx 1$, where the ``hint
of curvature'' in the redshift--distance relation is observed, our results fit the
observations better than the FRW $\Omega=1$ prediction. However, a still
better fit is provided by the FRW $\Omega =0.2$ curve.

Due to the possible (e.g. evolutionary) corrections to the redshift--distance
relation and the minute dissimilarities between FRW and LTB relations no
assertions as to the best fit to observations seem irrefutable. Nevertheless, it
is worth noticing that the model presented here supplies us with means to
reinterpret the correlation between the theory and observational results
(perhaps improving agreement between the two).

As a final element of the discussion of observational properties of our
model with a local LTB void, let us concentrate on the Hubble parameter
(constant) measurement. In a spherically symmetric model we have two
``Hubble parameters'': $H_{r} (t,r)$ for the local expansion rate in the radial
direction and $H_{\bot} (t,r)$ for expansion in the perpendicular direction.
Usual definitions give:
\begin{subequations} \label{hubble}
\begin{equation} \label{hubbler}
H_{r}=\frac{\dot{l}_{r}}{l_{r}}=\frac{\dot{R}^{\prime}}{R^{\prime
}},
\end{equation}
\begin{equation} \label{hubblep}
H_{\bot}=\frac{\dot{l}_{\bot}}{l_{\bot}}=\frac{\dot{R}}{R},
\end{equation}
\end{subequations}
where $l$ denotes the proper distance, i.e. $l_{r} = R^{\prime} (t,r) f^{-1}
dr$ and $l_{\bot} = R(t,r) d\Omega$. Due to the fact that there are both
gravitational and expansion redshifts contributing to the total $z$, neither of
the Hubble parameters $H_r$, $H_{\bot}$ is fully analogous to the FRW's
$H_{FRW}=\dot{a}/a$. The closest analogy exists for small separations of
the source and the observer. Let us notice that the integral in (\ref{grco})
can be rewritten as:
\[
 \int_{0}^{r_{1}} dr \frac{4 \pi \rho r}{{(1-a_{1})}^{2}},
\]
and for small $r_{1}$ can be neglected when compared to the first term of
(\ref{grco}). Expanding the logarithms on both sides we see that light
emitted at ($t_{e},r_{e}$) and observed at ($t_0,0$) satisfies for small $r_{e}$
or small ($t_0-t_{e}$):
\[
 z (t_{e},r_{e})=\dot{R} (t_{e},r_{e}),
\]
where $t_{e}$ is $T(r_{e})$ from (\ref{lightcone}) with the initial condition
$T(0)=t_0$. Using (\ref{hubblep}) and (\ref{lumdis}) we get for small
$r_{e}$:
\begin{equation}
 z (t_{e},r_{e})=H_{\bot}(t_{e},r_{e}) d_{L} (t_{e},r_{e}),
\end{equation}
which is formally analogous to the FRW result. Two main differences are
that our relation is {\em local} and that from cosmological observations (on
small scales) we obtain the angular Hubble parameter $H_{\bot}=\dot{R} /
R$ rather than $H_{FRW}=\dot{a}/a$.

In general, if we lived in a local LTB void and the $z$ vs. $d_L$ relation
differed from the FRW one as described in this paper, but we were biased
by our theoretical prejudice and interpreted cosmological observations
through an FRW model, we would expect the value of the Hubble
parameter to be position, or rather $d_L$, dependent.

To explore this possibility let us recall that in FRW cosmology the exact
result for the Hubble relation ($z$ versus $d_L$) in the matter dominated
universe is (\cosmoloc):
\begin{equation} \label{hubblefrw}
H_0 d_L ={q_0}^{-2} \left[ zq_0 + \left(q_0-1\right)\left(\sqrt{2zq_0+1} -1
\right) \right],
\end{equation}
where $q_0 \equiv -\ddot{a}(t_0)/a(t_0){H_0}^2$ is the deceleration
parameter.

\begin{figure}[ht]
\vspace{3.4in} \relax \noindent \relax
\includegraphics{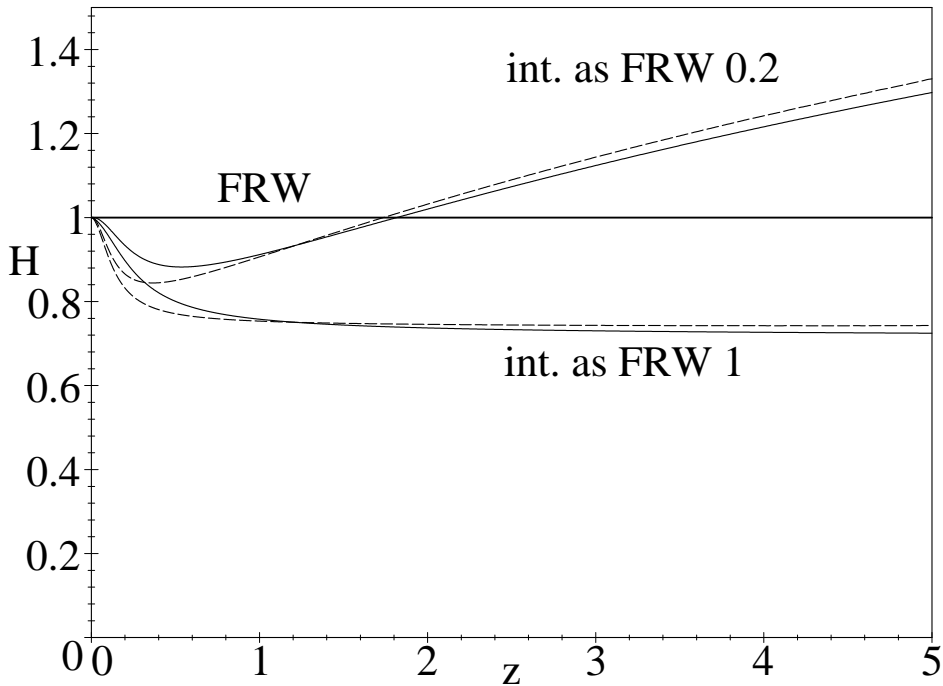}
\caption{{\tenrm\baselineskip=10pt The ``observed'' Hubble constant
$H$ (in units of the local measurement) as a function of the redshift $z$.
\label{f10}}}
\end{figure}

On cosmologically very small distances we measure the same value of
$H_0$ independently of the model (we call this value ``the local
measurement''). This stems from the fact that, due to our assumptions, very
close to the centre ($r \ll 1$) the model is well approximated by the FRW
universe with $\Omega=0.2$. Obviously, if the universe were locally LTB
rather than FRW, then the Hubble parameter based on the observed (LTB)
values of $z$ and $d_L$, but inferred through an FRW relation
(\ref{hubblefrw}), would be position (redshift) dependent. The dependence
of the Hubble parameter $H$ (in units of the $H_0$ value as measured
locally) on the redshift $z$ is shown in Figure \ref{f10}. For the sake of the
clarity of presentation, only the results for the ``younger'' case (I) are
included in the figure. Qualitatively, the ``older'' case (II) exhibits the same
behaviour. Quantitatively, as seen from the $z$ vs. $d_L$ relations of
Figure \ref{f7} and (\ref{hubblefrw}), the ``observed'' values of the Hubble
parameter (in the units of the local measurement) for the case (II) become
increasingly smaller than those of the case (I) ($\approx 5 \%$ smaller for
$z \approx 4$).

The $H(z)$ relation depicted in Figure \ref{f10} strongly depends on the
choice of the FRW case --the choice of the deceleration parameter $q_0$ in
(\ref{hubblefrw})-- through which we interpret the observations. The main
feature in the FRW $\Omega=1$ interpretation is a distinct monotonic
correlation between $H$ and $z$ (or $d_L$).  The values of the
``observed'' Hubble constant decrease with the redshift $z$ asymptotically
approaching the background limit. If we interpret the results within the
FRW $\Omega<1$ framework, the ``observed'' values of the Hubble
constant first decrease with $z$ and then asymptotically increase to some
background limit. (This asymptotic behaviour becomes clearly visible only
for very large $z$.) The position of the minimum in $H$ depends on the
size of the LTB void.  Neither of these patterns is evident in the reported
measurements of the Hubble constant.

However, one has to remember that the measurements of the Hubble
constant come inseparable from their own observational uncertainties
(estimates of the distance using a distance ladder, corrections for peculiar
motion etc.). Moreover, it is not the broader idea of a local LTB void
embedded in a globally FRW universe that fails here. A choice of density
distributions $\Omega(z)$ more complex than (\ref{densdistr}) could result
in a better agreement with observations.

We think that the variance in $H(z)$ exhibited in Figure \ref{f10} is a
valuable feature of the model. The ratios of the highest to the lowest value
of the ``observed'' $H$ within the range of redshifts of Figure \ref{f10} are
$\approx 1.5$. At the same time, the values of $H_0$ reported to date
span the range $40$ to $100 \mbox{ km } {\mbox{s}}^{-1}
{\mbox{Mpc}^{-1}}$ (with standard errors quoted frequently as $10 \mbox{
km } {\mbox{s}}^{-1} {\mbox{Mpc}^{-1}}$ or less!). Inhomogeneities similar
to the LTB void presented here might provide an explanation for this.

\section{DISCUSSION} \label{discuss}

The LTB voids presented here decrease their density contrasts (the depth of
the void with respect to the FRW background) when evolved back in time.
At early times they are almost homogenized: at $t/(t_0+\beta_0) \simeq
10^{-6}$ we have $|{\rho}_{LTB}(r)/{\rho}_{FRW}-1| \simeq 10^{-5}$
everywhere. This corresponds to a universe which at the beginning is very
similar to the FRW one, but different at late stages.

In a model utilizing a local LTB void, while retaining the accomplishments
of the FRW cosmology in dealing with epochs preceding the matter
dominated era, we can gain new freedom in modelling the more recent
universe. We can solve the age of the universe problem by assuming
$\beta(r) = const \neq 0$, provide the excess power observed on scales of
5---10,000 km s$^{-1}$ in modelling structure formation (see \motat) and
provide an explanation for the wide range of reported values of the Hubble
constant.

The use of a locally inhomogeneous model enriches the spectrum of
available interpretations. The ultimate question to be addressed, however, is
the agreement between predictions of the model and observations. As has
been demonstrated by the results of the present work, the observationally
based density distributions spanning the range of $\Omega \in (0.2;1)$
satisfy the observational tests. The agreement with observations is achieved
both in the case of the model with the ``age of the universe''  equal to that
of the critical ($\Omega_0 = 1$) FRW one and the ``older'' case with
$t_0+\beta_0$ equal to the FRW $\Omega_0=0.2$ value. Theoretical
prejudice (as powerful as the inflationary paradigm) might favour the critical
FRW model.

\acknowledgements

The authors thank N. Kaiser, A. Krasi\'{n}ski and S. J. Lilly for discussions.
This work was supported by the Natural Sciences and Engineering Research
Council of Canada. One of the authors (DCT) thanks the Ministry of
Colleges and Universities of the Province of Ontario for financial support
through an Ontario Graduate Scholarship.

\end{document}